# Beyond Autophagy: VPS39 Deficiency Triggers Migrasome-Driven Stress Adaptation Revealed by Super-Resolution Imaging

Xuelei Pang,[1†] Weiyun Sun,[2, 3†] Ning Jing[1], Wenwen Gong,[2, 3] Cuifang Kuang,[2] Xu Liu,[2] Hanbing Li,[3]* Yu-Hui Zhang,[1]* Yubing Han[1,2]*

[1] MOE Key Laboratory for Biomedical Photonics, Advanced Biomedical Imaging Facility−Wuhan National Laboratory for Optoelectronics, Huazhong University of Science and Technology, Wuhan, Hubei 430070, China

[2] State Key Laboratory of Extreme Photonics and Instrumentation, College of Optical Science and Engineering, Zhejiang University, Hangzhou, China

[3] Institute of Pharmacology, College of Pharmaceutical Sciences, Zhejiang University of Technology, Hangzhou, China

[†]These authors contributed equally

*Corresponding Authors：zhangyh@mail.hust.edu.cn (Yu-Hui Zhang); hanniballee@zjut.edu.cn (Hanbing Li); hanyubing@zju.edu.cn (Yubing Han)

**Abstract**

Autophagy and migrasome formation constitute critical cellular mechanisms for maintaining cellular homeostasis, however, their potential compensatory interplay remains poorly understood. In this study, we identify VPS39, a core component of the HOPS complex, as a molecular switch coordinating these processes. Genetic ablation of VPS39 not only impairs autophagic flux but also triggers cell migration through RhoA/Rac1 GTPases upregulation, consequently facilitating migrasome formation. Using super-resolution microscopy, we further demonstrate that migrasomes serve as an alternative disposal route for damaged mitochondria during VPS39-induced autophagy impairment, revealing a novel stress adaptation mechanism. Our work establishes a previously unrecognized autophagy-migrasome axis and provides direct visual evidence of organelle quality control via migrasomal extrusion. These findings position VPS39-regulated pathway switching as a potential therapeutic strategy for neurodegenerative diseases characterized by autophagy dysfunction.

**Introduction**

Autophagy, a sophisticated cellular process, plays a pivotal role in the degradation and recycling of intracellular components, thereby ensuring cellular homeostasis[1-3]. Central to this mechanism is the biogenesis of the autophagosome, a specialized structure that fuses with lysosomes to facilitate the breakdown and reuse of cellular materials[4-6]. This process is orchestrated by a series of autophagy-related genes (ATG), with ATG7 serving as a crucial enzyme in multiple stages of autophagosome formation[7]. Widely recognized autophagy markers such as LC3-II and P62 offer insights into this process, where LC3-II participates directly in autophagosome formation and P62 acts as a selective autophagy receptor[8, 9]. Inhibition of lysosomal fusion with autophagosomes leads to blocked autophagic flux, resulting in an accumulation of undegraded autophagosomes, potentially contributing to various neurodegenerative diseases[10, 11]. VPS39, a pivotal element of the homotypic fusion and protein sorting (HOPS) complex, is indispensable for the maintenance of lysosomal function and integrity[12, 13]. The suppression of VPS39 expression disrupts autophagic flux in human myoblasts, culminating in defects in muscle differentiation[14].

To counteract stress induced by autophagy dysfunction, cells activate compensatory mechanisms to restore balance[15]. Increasing research underscores the complex interplay between autophagy and cell migration, involving intracellular homeostasis, signal transduction, and cytoskeletal reorganization[16-18]. In breast cancer, autophagy suppresses cell migration, maintaining homeostasis and preventing mutations[19, 20]. Conversely, in immune cells like T cells, autophagy is downregulated to prioritize rapid migration to sites of infection or inflammation[21, 22]. Rho family GTPases, particularly RhoA and Rac1, are critical in these processes, orchestrating cytoskeletal dynamics and cell movement to balance autophagy and migration[23-25].

Migrasomes are recently unveiled cellular structures that form during cell migration, guiding cells along pathways that optimize survival and functionality[26-28]. Both migrasomes and autophagy are pivotal in regulating the internal and external cellular environments[29]. Migrasomes contribute to migration, waste management, and stress mitigation[30, 31]. Increased migration enhances retraction fibers, facilitating migrasome formation, whereas reduced mobility constrains this process[30, 32, 33]. Involved in various physiological and pathological contexts, migrasome production ramps up under stress, boosting migratory capabilities and aiding the clearance of damaged components, thereby sustaining cellular homeostasis[34]. Research by Yu Li et al. demonstrates that impaired mitochondria are expelled via migrasomes, preserving cellular energy balance and metabolic health[35, 36].

Here, we report that VPS39 deficiency mitigates cellular stress induced by autophagy inhibition through enhanced migrasome generation. Mechanistically, the loss of VPS39 suppresses autophagic flux and enhances cell migration by upregulating the expression of RhoA and Rac1, thereby regulating migrasome formation. This discovery provides new insights into cellular stress responses, migrasome dynamics, and potential therapeutic applications.

**Result**

**VPS39 Knockdown Promotes Migrasome Biogenesis**

We set out to uncover the ultrastructure of migrasomes using super-resolution microscopy. The long-term structured illumination microscopy (SIM) images showed the dynamic biogenesis of migrasomes in TSPAN4-EGFP-expressing L929 cells (Fig. 1a). Initially observed at the intersections of retraction fibers, migrasomes expanded in size and translocated along these fibers during the 5-h observation period. This spatial specificity suggests migrasomes preferentially form at retraction fiber crosspoints, with their subsequent growth and migration dependent on these structures serving as physical scaffolds[28, 37]. Notably, stimulated emission depletion (STED) microscopy compared to confocal microscopy, revealed that retraction fibers exhibit a tubular lumen (Fig. S1). Line analysis shows that the lumen width of the retraction fibers is 0.27 μm (Fig. S2 a-c), potentially pivotal for the mechanics of migrasome formation.

Meanwhile, SIM-based colocalization studies revealed that some actin filament bundles at the cellular periphery were mainly enwrapped by migrasomes, as observed in both live-cell and fixed-cell preparations (Fig. 1b,c; Fig. S3,4), underscoring the crucial role of microfilaments in orchestrating migrasome formation and trafficking dynamics[29, 30, 38]. While as a contrast, the signal resolution was insufficient to delineate precise spatial relationships between migrasomes and microfilaments in confocal images (Fig. S5).

We then established stable L929 cell lines with VPS39 overexpression or knockdown (Fig. 1d; Fig. S6a, b) and focused on shVPS39-2, which achieved the most efficient VPS39 depletion (71% reduction). VPS39 showed strong colocalization with lysosomes (Pearson's r = 0.75) and moderate association with Rab7 (r = 0.54) in confocal imaging analysis (Fig. S7a). Interestingly, we observed a significant increase in migrasome formation in VPS39-knockout cells, a phenomenon consistently detected in both L929 and U2OS cell lines in SIM imaging (Fig. 1e, f). We introduced fibronectin (FN; a promoter of cell adhesion and migration[39]) and SAR407899 (SAR; the ROCK1 inhibitor and inhibits migrasome biogenesis[40]) to further examine the regulatory factors of migrasome biogenesis (Fig. 1g). The SIM images and relative quantitative statistical results show that the number of migrasomes significantly increased after VPS39 knockdown or FN treatment while it sharply decreased after SAR treatment (Fig. 1g, h). These above results indicate that VPS39 deficiency promotes migrasome biogenesis, probably by enhancing cellular migration.

Previous studies have demonstrated that Reactive Oxygen Species (ROS) serve as pivotal signaling molecules in cell migration[41]. We then assessed intracellular ROS levels by determine 2',7'-dichlorodihydrofluorescein diacetate (DCFDA, a classic fluorescent probe for detecting intracellular ROS[42]) fluorescence intensity (Fig. 1i; Fig. S8). The results show that in VPS39-deficient cells, ROS levels were significantly elevated, and this phenotype was rescued by VPS39 overexpression. Meanwhile, pharmacological inhibition of migrasome formation using SAR led to a dramatic increase in intracellular ROS, whereas FN-mediated stimulation of migrasome generation partially alleviated oxidative stress. Collectively, these findings reveal that VPS39 deficiency induces oxidative stress while enhancing migrasome formation, suggesting migrasomes may serve as a compensatory mechanism to alleviate cellular ROS accumulation.

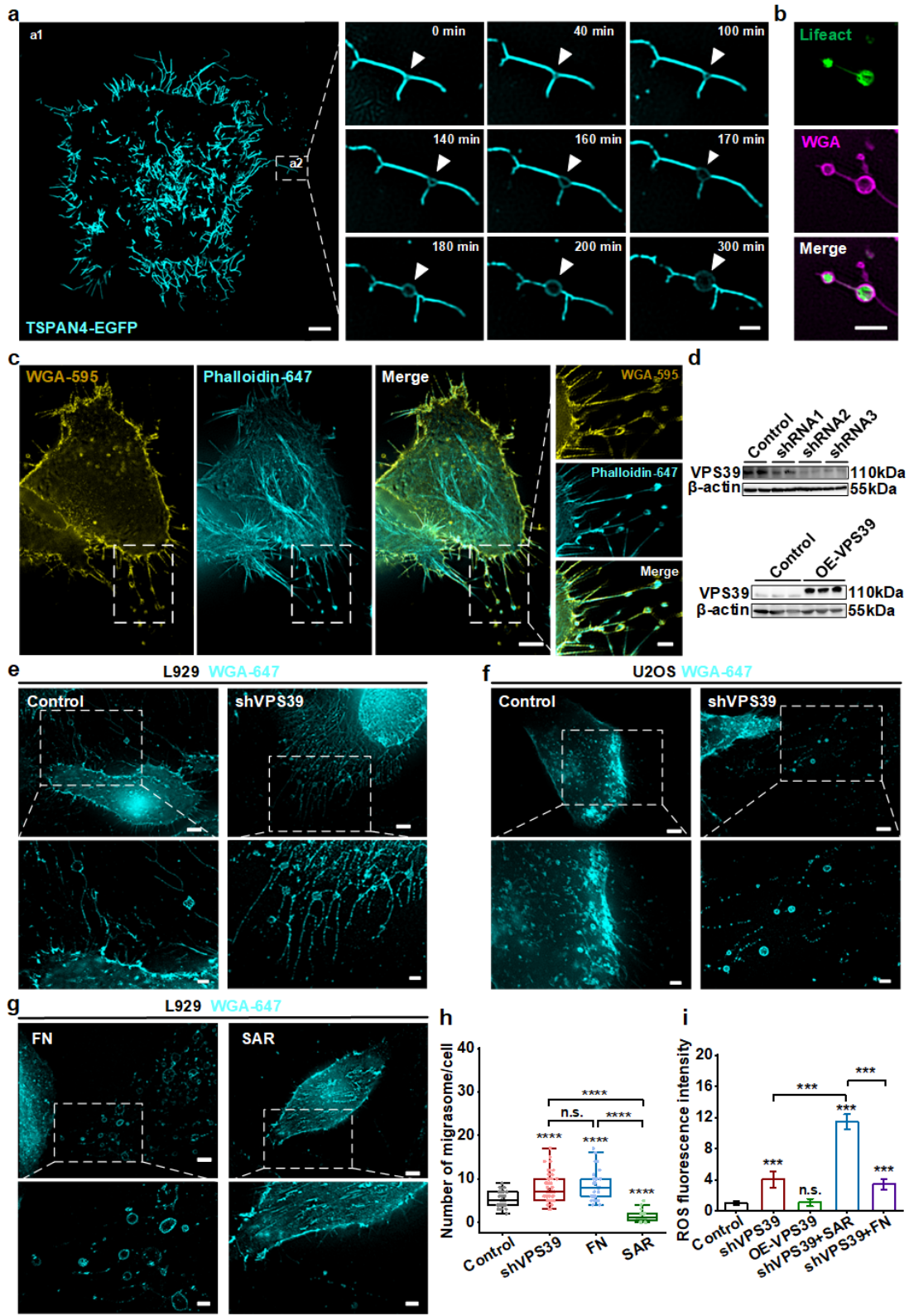

**Fig1. VPS39 Knockdown Increases Migrasome Formation. a** SIM images (10-min intervals) of L929 cells expressing the migrasome marker TSPAN4-EGFP. Time-lapse magnification of region of interest (ROI) captures sequential migrasome biogenesis along retraction fibers. **b, c** SIM imaging of migrasomes (WGA-647, magenta; WGA-595, yellow, pre-fixation) and actin filaments (Lifeact-

EGFP, green; Phalloidin-647, cyan, post-fixation) in (b) live or (c) fixed L929 cells. **d** Western blot (WB) analysis of VPS39 protein expression in control L929 cells, VPS39-knockdown cells (expressing three distinct shRNAs), and VPS39-mCherry-overexpressing cells (OE-VPS39). β-Actin: loading control. **e, f** Representative SIM images of migrasomes (WGA-647) in (e) L929 and (f) U2OS cells under Control or VPS39-knockdown (shVPS39) conditions. Bottom panels: Magnified ROIs. **g** Representative SIM images of migrasomes in L929 cells with FN stimulation and SAR inhibition. Bottom: magnified ROIs. **h** Quantification of migrasome number per cell under indicated conditions using SIM imaging. Data was acquired from n ≥ 24 cells across 9 SIM images per condition. **i** Intracellular ROS levels (mean DCFDA fluorescence intensity) in L929 cells under indicated conditions using confocal imaging. Data in (h, i) represent mean ± SEM; ***$p < 0.001$, ****$p < 0.0001$, n.s. not significant (two-tailed unpaired t-test). Scale bars: 5 μm (a1, c, e-g); 2 μm (a2, b, c, e-g: magnified panels).

**Impaired Autophagic Flux and Lysosomal Function in VPS39 Deficiency**

A tandem pMRX-IP-GFP-LC3-RFP-LC3ΔG plasmid construct was then employed to investigate the role of VPS39 in autophagy regulation[43]. Since GFP is quenched in the acidic (auto)lysosomal environment, monitoring the RFP/GFP signal ratio provides a quantitative measure of autophagic flux. The confocal imaging results revealed increased RFP signals upon starvation in control cells, while chloroquine (CQ) treatment or VPS39 knockdown resulted in abnormal GFP retention (Fig. 2a; Fig. S9, S10). Quantitative analysis confirmed starvation enhanced autophagic flux (2.3±0.2; $p<0.01$), whereas VPS39 depletion suppressed basal autophagic flux by 67% (0.5±0.3; $p<0.001$) and abolished starvation-induced autophagic response (0.6±0.15 vs 2.3±0.2; $p<0.001$) (Fig. 2b).

We next examined the expression profiles of key autophagy- and lysosome-associated proteins to elucidate the functional consequences of VPS39 depletion. Western blot (WB) analysis revealed that VPS39 depletion significantly elevated autophagic substrates p62 and LC3-II, with the most pronounced accumulation observed under starvation combined with CQ treatment (Fig. 2c-e). Immunofluorescence confirmed aberrant LC3-II puncta and p62 signals aggregates in VPS39-deficient cells under both basal and starved conditions (Fig. 2f, g; Fig. S11 a, b), consistent with impaired autophagosome-lysosome fusion[12, 44]. Meanwhile, VPS39 knockdown selectively increased p62 and LC3-II without altering Lamp1/Lamp2 protein levels (Fig. 2h, i), indicating specific disruption of autophagic flux rather than

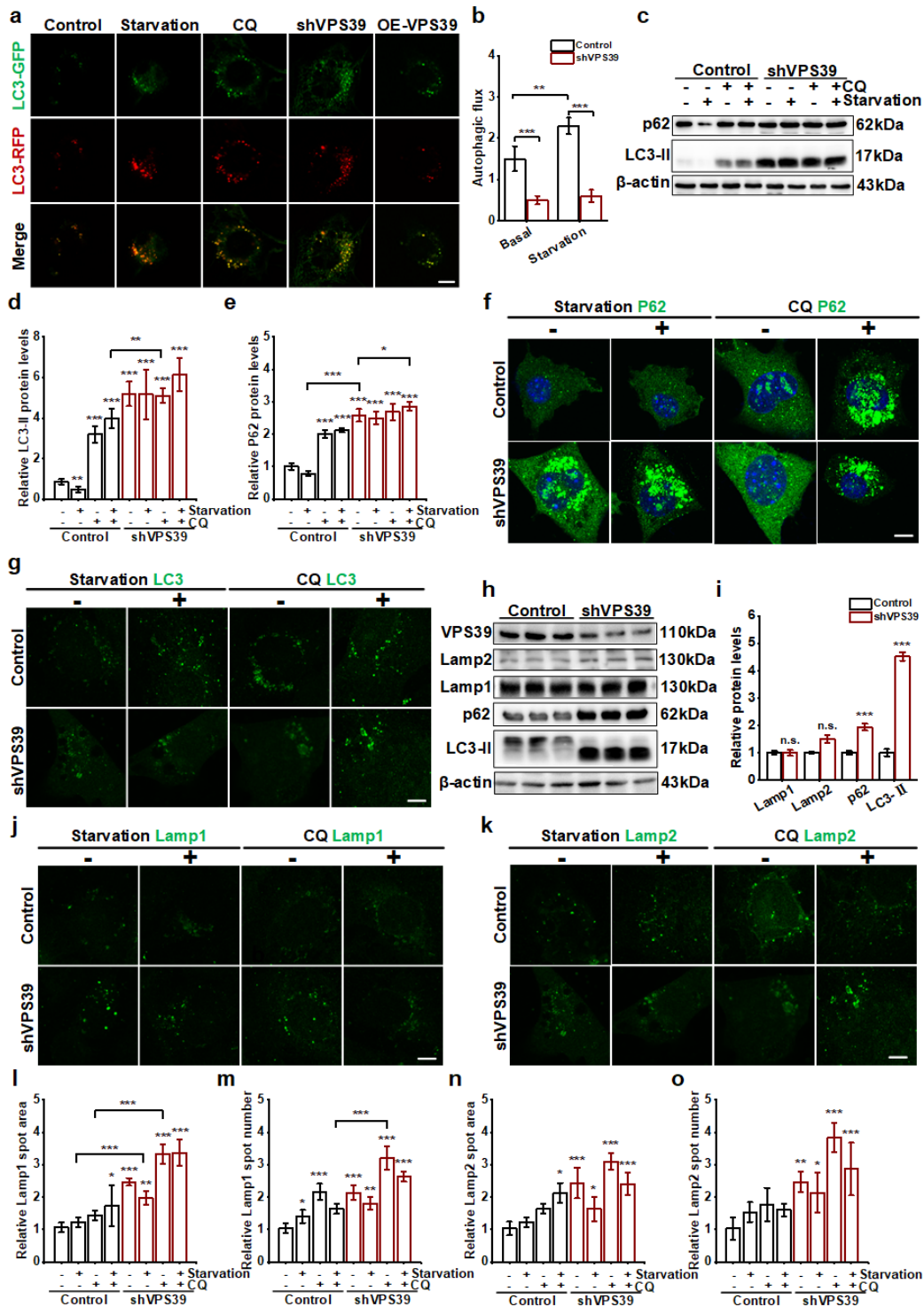

**Fig2 VPS39 depletion impairs autophagic flux and lysosomal function. a** Confocal images of L929 cells expressing the autophagic flux sensor GFP-LC3-RFP-LC3ΔG, showing autophagosomes (yellow puncta) and autolysosomes (red puncta) under control, starvation, CQ, shVPS39, and OE-VPS39. **b** Quantification of autophagic flux (RFP/GFP intensity ratio) in control and shVPS39 L929 cells under basal and starvation conditions from confocal images. **c** WB analysis of p62 and LC3-II expression in control and shVPS39 L929 cells under the indicated conditions of starvation and chloroquine (CQ) treatment. β-actin served: loading control. **d, e** Quantification of relative p62 (d) and LC3-II (e) protein levels under the indicated conditions of starvation and CQ treatment. **f, g** Endogenous immunofluorescence detection of autophagy-related proteins, p62 (f) and LC3-II (g), in control and shVPS39 L929 cells. **h** WB analysis of VPS39, Lamp2, Lamp1, p62, and LC3-II expression in control and shVPS39 L929 cells. β-actin: loading control. **i** Quantification of protein levels from (h). **j, k** Endogenous immunofluorescence detection of Lamp1 (j) and Lamp2 (k) in control and shVPS39 cells under Basal, Starvation, or CQ conditions. **l, m** Quantification of the area and number of Lamp1-positive vesicle regions in L929 cells under the indicated conditions of starvation and CQ treatment. **n, o** Quantification of the area and number of Lamp2-positive vesicle regions in L929 cells under the indicated conditions of starvation and CQ treatment. Data in (d, e, i, l-o) normalized to control, data in (b, d, e, i, l-o) represent mean ± SEM, *$p < 0.05$, **$p < 0.01$, ***$p < 0.001$, n.s. not significant ($p > 0.05$). Scale bars: 10 μm (a, f, g, j, k).

lysosomal biogenesis defects. Although total Lamp1/2 protein levels remain unchanged, endogenous immunofluorescence were observed for Lamp1/2 positive structures, with both lysosomal area ($p<0.001$) and abundance ($p<0.01$) elevated (Fig. 2j-o). In contrast, Lyso Tracker staining showed a reduction in the number of acidic compartments ($p<0.01$), despite their increased area ($p<0.001$, Fig. S12a-c). This paradox reveals that VPS39 knockdown impairs HOPS-mediated autophagosome-lysosome tethering, stalling lysosomal vesicles at intermediate stages. Undigested cargo accumulation causes osmotic swelling, while failed ATPase recruitment induces luminal alkalinization—explaining Lyso Tracker signal reduction despite structural proliferation and exacerbating cellular stress.

**Migrasome Biogenesis Upregulated upon Autophagy Blockade**

To further investigate the mechanism by which autophagy flux regulates migrasome formation, we first performed scratch wound healing assays. The imaging and quantitative results reveal that treatment with autophagy inhibitor CQ significantly enhanced cell migration, whereas autophagy inducer rapamycin (RAPA, mTORC1 signaling pathway inhibitor capable of inducing autophagy[45]) markedly suppressed migration (Fig. 3a, b), suggesting an inverse relationship between autophagy activity and migratory capacity.

Given the inverse relationship between autophagy inhibition and cell migration, we next investigated whether autophagy status similarly impacts migrasome formation. We knocked down ATG7, a crucial autophagy-related gene, and the WB analysis of

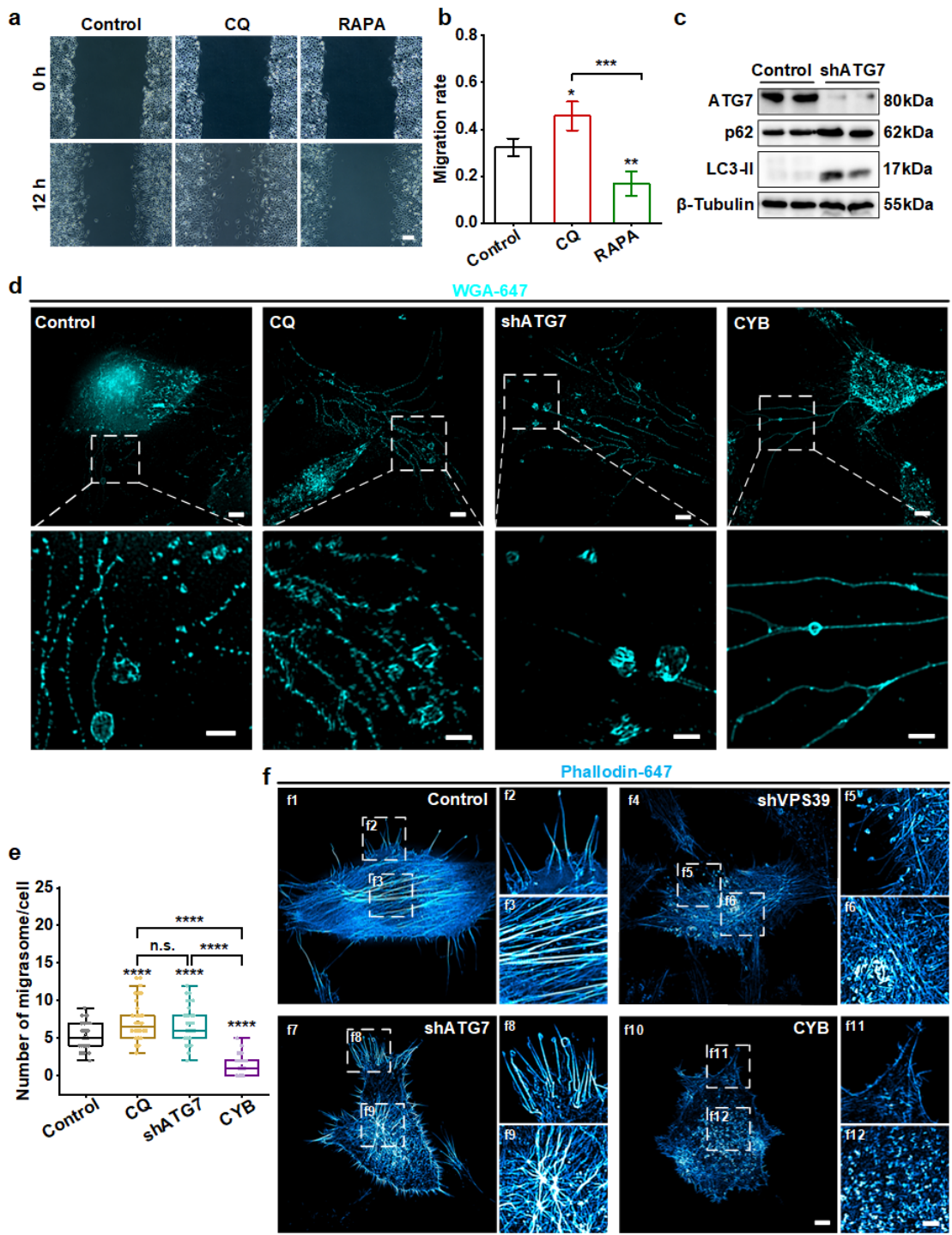

**Fig3 Promoted cell migration after inhibiting autophagy. a** Widefield images from scratch wound healing assay of control, CQ, and RAPA-treated cells at 0 h and 12 h post-scratching. **b** Quantification of wound closure percentage at 12 h of the control, CQ, and RAPA treated cells. **c** WB analysis of ATG7, p62, and LC3-II protein expression in control and ATG7-knockdown (shATG) cells. β-actin: loading control. **d** Representative SIM images of migrasomes (WGA-647) in control, CQ, ATG7-knockdown (shATG7), and CYB-treated L929 cells. Bottom: Magnified ROIs. **e** Quantification of migrasome number per cell under indicated conditions, corresponding to representative images in d. **f** Representative SIM images of actin filaments (Phalloidin-647) in control, shVPS39, shATG7, and CYB-treated L929 cells. **f1-f8**: magnified ROIs. Data in (b, e) represent mean ± SEM; n = 3 independent experiments. *P < 0.05, **P < 0.01, ***p < 0.001, ****p < 0.0001, n.s. not significant (two-tailed unpaired t-test). Scale bars: 100 μm (a); 5 μm (d, f); 2 μm (d, f: magnified panels).

autophagy markers confirmed that ATG7 knockdown significantly inhibited autophagy flux, as evidenced by increased P62 accumulation and reduced LC3-II levels (Fig. 3c, Fig. S11). Super-resolution SIM imaging reveals differential migrasome distribution patterns under varying treatment conditions (Fig. 3d), and quantitative analysis demonstrated that autophagy inhibition, whether pharmacologically induced (CQ) or genetically mediated (shATG7), significantly enhanced migrasome abundance compared to control cells (p < 0.0001, Fig. 3e). Conversely, treatment with cytochalasin B (CYB, a cytoskeletal depolymerizing agent targeting actin polymerization[46]), a selective inhibitor of actin polymerization, significantly attenuated migrasome abundance (P < 0.001, Fig. 3e), establishing a direct functional relationship between cytoskeletal dynamics, migrasome biogenesis, and cellular migration capacity. Collectively, these data demonstrate a mechanistic interplay between autophagic activity and migrasome formation, suggesting that autophagy regulates cell motility in part through modulation of migrasome-dependent processes.

To further characterize the cytoskeletal changes associated with autophagy inhibition, we conducted SIM imaging of F-actin to examine its detailed morphology across differential experimental conditions (Fig. 3f). In control group, cells exhibited well-defined cortical actin networks with moderate stress fiber formation (Fig. 3f, top left panels). As a striking contrast, cells depleted of either VPS39 or ATG7 displayed robust cytoskeletal alterations. We observed enhanced peripheral stress fibers bundles formation, pronounced actin reorganization, and a proliferation of actin-rich protrusive structures like membrane ruffles and lamellipodia (Fig. 3f). These structures are hotspots for cell migration and provide the foundation for increased migrasome formation[33, 47]. Furthermore, this cytoskeletal remodeling, resulting from autophagy deficiency, was accompanied by a notable increase in migrasome abundance (Fig. 1h). Conversely, CYB treatment induced substantial actin depolymerization, characterized by marked reduction in stress fibers and diminished F-actin fluorescence intensity throughout the cytoplasm (Fig. 3f, bottom right panels). These findings suggest that autophagy deficiency, whether induced by VPS39 or ATG7 silencing, promotes actin cytoskeletal rearrangements that enhanced cell motility and favor migrasome formation.

**Autophagy Impairment by VPS39 Loss Triggers RhoA/Rac1-Mediated Migration**

To determine whether VPS39 knockdown affects cell migration and to identify the underlying molecular mechanisms, we performed scratch wound assays and analyzed cytoskeletal regulatory pathways. Wound healing assays revealed that VPS39-depleted cells exhibited significantly enhanced migration, with wound closure rates 1.8-fold (12h) and 1.4-fold (24h) higher than controls ($p<0.01$). Conversely, VPS39 overexpression yielded a slight, non-significant reduction in migration speed (Fig. 4a, b). These data indicate that VPS39 deficiency promotes cell migration, consistent with the results of CQ treatment (Fig. 3a, b), suggesting that autophagy inhibition can enhance cellular motility.

Given the inverse relationship between VPS39-dependent autophagy and cellular motility, we investigated whether autophagy inhibition selectively modulates Rho GTPase activity. We focused on three canonical Rho GTPase family members: RhoA (stress fiber formation), Rac1 (lamellipodia formation), and CDC42 (filopodia formation and directional persistence). Protein expression analysis demonstrated that knockdown of the essential autophagy protein ATG7 elevated RhoA and Rac1 expression (Fig. S14a). Similarly, VPS39 knockdown markedly upregulated levels of these two Rho GTPase without affecting CDC42 expression (Fig. 4c, Fig. S14b), indicating that autophagy inhibition selectively regulates the activation of cytoskeletal remodeling programs, thereby driving enhanced cellular migration.

To systematically evaluate how autophagy regulates these migration-related proteins, we conducted immunofluorescence analysis under conditions of impaired autophagy (VPS39 knockdown, CQ treatment) and enhanced autophagy (VPS39 overexpression, RAPA treatment), examining their endogenous expression (Fig. 4d-g). The results confirmed that autophagy inhibition (via either VPS39 knockdown or CQ treatment) significantly increased Rac1 and RhoA expression (Fig. 4d, e, g), while RAPA-induced autophagy markedly reduced their fluorescence intensity (Fig. 4g). Notably, CDC42 expression remained stable across all experimental conditions (Fig. 4f, g). These findings demonstrate that autophagy inhibition selectively upregulates specific Rho GTPases (Rac1 and RhoA) while leaving others (CDC42) unchanged. The differential regulation suggests a compensatory mechanism whereby cells enhance

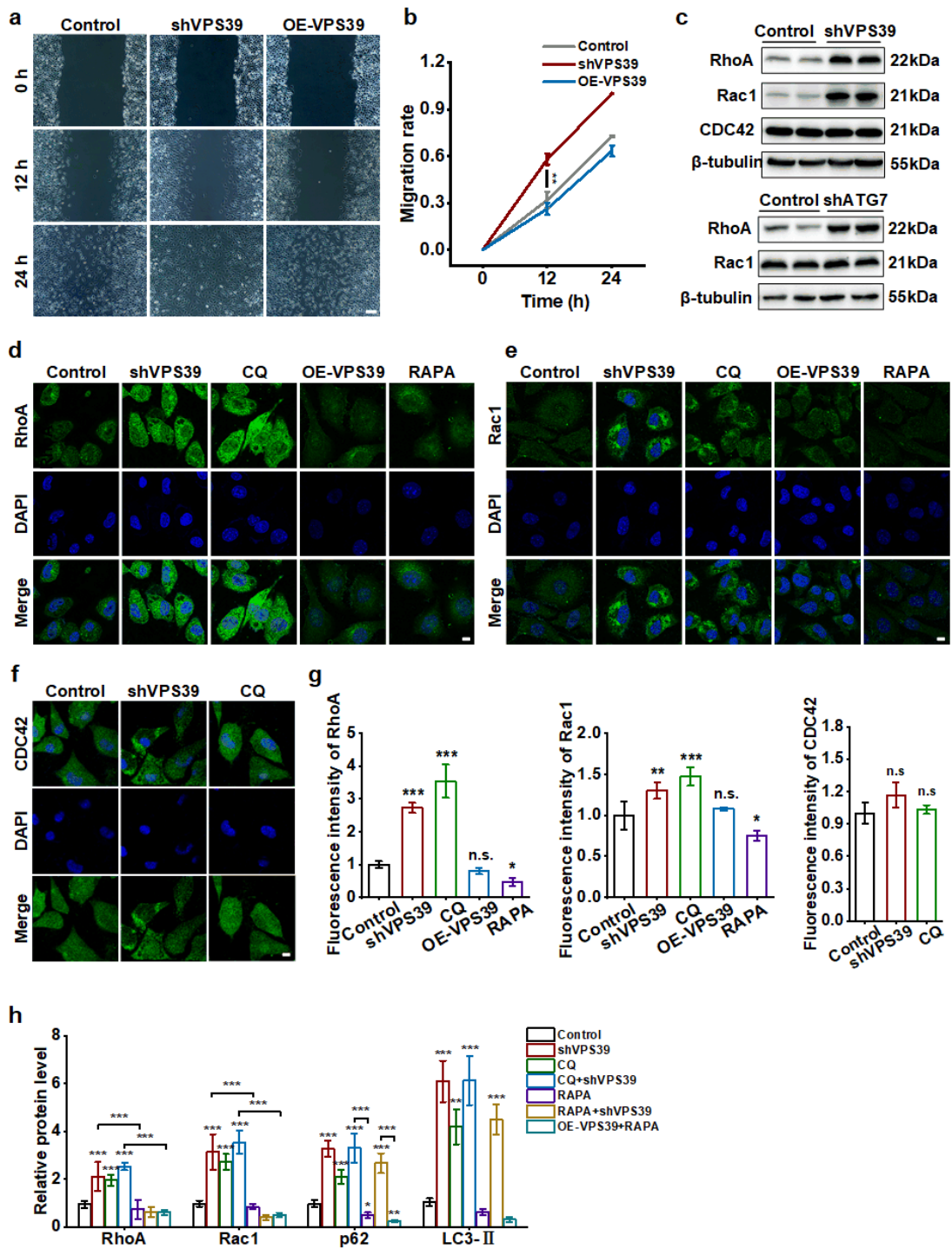

**Fig4 Enhanced expression of cell motility factors promotes migrasome formation. a** Widefield images from scratch wound healing assay showing wound closure in control, VPS39-knockdown, and VPS39-overexpression groups at 0 h, 12 h, and 24 h after wounding. **b** Quantification of wound closure percentage at 12 h and 24 h of the control, VPS39-knockdown, and VPS39-overexpression cells. **c** WB analysis of RhoA, Rac1, and CDC42 in control, VPS39-knockdown, and ATG7-knockdown cells. β-tubulin: loading control. **d-f** Confocal immunofluorescence (IF) images of endogenous (d) RhoA, (e)Rac1, and (f) CDC42 in control, shVPS39, CQ-treated, OE-VPS39, and RAPA-treated cells. CDC42 was examined only in control and shVPS39 groups. Nuclei: DAPI (blue). **g** Quantification of RhoA, Rac1, and CDC42 IF intensity under specified conditions. **h** WB analysis of cytokines (RhoA, Rac1) and autophagy markers (P62 and LC3-II) across multiple treatment conditions. Scale bars: 100 μm (a); 10 μm (d, e, g).

specific migratory pathways during autophagy impairment while preserving CDC42-dependent functions such as cell polarity. This selective modulation of cytoskeletal regulators explains the molecular mechanism underlying enhanced cell migration during autophagy deficiency[48].

We next investigated the relationship between autophagic status and migration-related proteins across multiple experimental conditions (Fig. 4h; Fig. S15). VPS39 knockdown effectively inhibited autophagic flux, evidenced by persistent accumulation of p62 and LC3-II. Notably, RAPA treatment could not rescue the impaired autophagic flux in VPS39-depleted cells, demonstrating that VPS39 deficiency causes an irreparable block in autophagosome-lysosome fusion that cannot be bypassed by upstream autophagy induction, confirming the essential role of VPS39 in regulating autophagic flux.

**VPS39 Loss Promotes Damaged Mitochondria Clearance via Migrasomes**

As a major site of ROS production, mitochondrial function is crucial for maintaining cellular homeostasis[49]. Given that VPS39 knockdown impairs autophagy and increases cellular ROS stress, we further investigated mitochondrial function using the JC-1 probe (Fig. 5a, b). JC-1 exhibits red fluorescence when aggregated in mitochondria with intact membrane potential ($\Delta\Psi m$), but shifts to green fluorescence as monomers in the cytoplasm when $\Delta\Psi m$ collapses[50]. The results showed that both control cells and VPS39-overexpressing cells maintained a high $\Delta\Psi m$. In contrast, VPS39-knockout cells exhibited a significant increase in green fluorescence, indicating mitochondrial membrane potential dysfunction. When additional interventions were applied to VPS39-deficient cells, although FN could enhance cell migration, it failed to restore depolarized $\Delta\Psi m$; Furthermore, the SAR treatment, which inhibits cell migration, significantly aggravated mitochondrial depolarization (Fig. 5a, b). Overall, VPS39 deficiency induces mitochondrial depolarization, which cannot be rescued by FN, and is exacerbated by SAR, indicating that the loss of VPS39 disrupts $\Delta\Psi m$ and is irreversible.

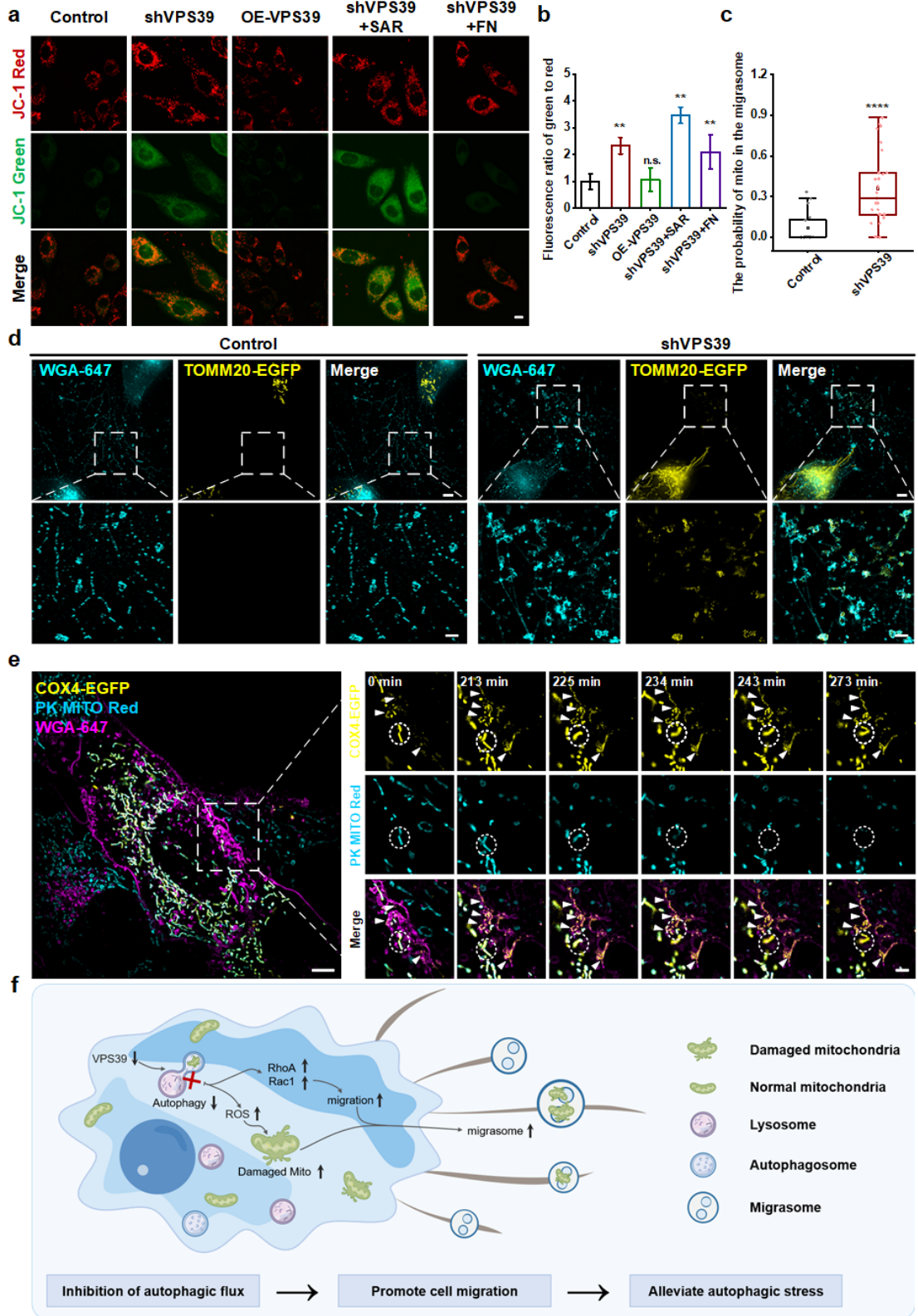

**Fig5 Extracellular release of damaged mitochondria to alleviate intracellular pressure. a** Confocal images and **b** quantitative analysis of the green/red fluorescence intensity ratio of JC-1 staining in L929 cells to assess mitochondrial membrane potential under indicated conditions. **c** Quantification of mitochondrial incorporation frequency in migrasomes of control and VPS39-

knockdown L929 cells. **d** Representative SIM images of migrasomes (WGA-647) and mitochondrial (TOMM20-EGFP) in control and shVPS39 L929 cells. Bottom: Magnified ROIs. **e** Encapsulation of damaged mitochondria by retraction fibers in VPS39-knockdown U2OS cells. Time-lapse SIM images (3-min intervals) of mitochondria (COX4-EGFP, green; PK MITO Red, blue) and migrasomes (WGA-647, magenta). **f** Molecular diagram of how cells alleviate autophagic inhibition pressure through migrasomes after VPS39 knockdown. Data in (b, c) represent mean ± SEM; n = 3 independent experiments. **$p < 0.01$, ***$p < 0.001$, ****$p < 0.0001$, n.s. not significant (two-tailed unpaired t-test). Scale bars: 10 μm(a); 5 μm (d, e,); 2 μm (d, e: magnified panels).

Notably, SIM images revealed a striking difference in mitochondrial distribution within migrasomes between control and VPS39-depleted cells (Fig. 5c, d). In control cells, mitochondria were observed in only 7% of migrasomes, whereas VPS39 deficiency led to a striking 5.1-fold increase, with mitochondria now present in 36% of these structures. These findings suggest that migrasomes may function as alternative compartments for mitochondrial segregation when conventional autophagy pathways are compromised. These observations were further corroborated by nearly 5 h of time-lapse SIM imaging experiments (Fig. 5e; Fig. S16). We employed a multi-label strategy: COX4-EGFP to label the overall mitochondrial structure, PK MITO Red as a $\Delta\Psi m$-dependent probe, and WGA-647 to label cell migrasomes and retraction fiber structures in VPS39-depleted U2OS cells (Fig. 5e). The time-lapse SIM images demonstrated the temporal dynamics of mitochondrial damage: progressive $\Delta\Psi m$ dissipation (indicated by PK MITO Red loss) while COX4-EGFP retention confirmed organelle identity. Remarkably, these mitochondria with dissipated membrane potential are subsequently encapsulated by WGA-647-labeled retraction fibers and systematically transported to and accumulated at the peripheral region of the cell (indicated with white arrowheads and dashed circles in Fig. 5e). Interestingly, we also observed an entire process of mitochondrial intercellular transfer via retraction fibers in control cells in long-term SIM imaging (Fig. S16). Originating at the cell periphery, a mitochondrion navigated through retraction fibers into a migrasome, underwent fusion with resident mitochondria, and was collectively transported to an adjacent cell within a 60-min window.

These above results demonstrate migrasome-dependent removal of damaged mitochondria[35]. Building on these findings, we propose that VPS39 deficiency triggers oxidative stress through defective autophagy, forcing cells to adopt migrasome-mediated mitochondrial expulsion as an essential homeostasis-preserving mechanism (Fig. 5f).

**Discussion and Conclusion**

Our study elucidates a pivotal regulatory axis connecting VPS39-mediated autophagy-lysosomal fusion, mitochondrial homeostasis, and migrasome biogenesis, revealing novel cellular adaptations to autophagic disruption. Key findings demonstrate that super-resolution microscopy (SIM/STED) captured migrasome genesis at retraction fiber intersections, with luminal structures suggesting a structural scaffold for biogenesis (Fig. 1a, Fig. S1, 2). SIM imaging reveals actin filaments encapsulated within migrasomes (Fig. S3, 4). This finding, combined with structural evidence of membrane expansion sites, strongly suggests that actin filaments constitute mechanism for migrasome biogenesis[26, 30, 37].

Consistent with classical theory, this study confirms that VPS39, as a core component of the HOPS complex, plays an essential regulatory role in autophagosome-lysosome fusion, with its absence directly disrupting the HOPS-mediated membrane fusion prohpess. Our finding reveals that VPS39 depletion results in the accumulation of damaged mitochondria, indicated by a 4-fold increase in ROS and a 43% decrease in $\Delta\Psi m$, due to impaired autophagic clearance. In this situation, cells activate a compensatory migrasome pathway—migration speed increases by 1.8 times and migrasome generation by 1.4 times ($p<0.0001$), alleviating the strain of autophagic inhibition by expelling damaged mitochondria.

Contrary to traditional views that autophagic inhibition generally suppresses migration[51], this phenomenon arises from the dual-edge effect of damage signals: accumulated ROS activate the Rac1/CDC42 pathway (functional activation), and our results demonstrated that VPS39 deletion specifically upregulates RhoA and Rac1 expression (transcriptional regulation) while CDC42 expression remains unchanged. This finding reveals a selective regulatory strategy that fine-tunes migration without compromising core functions like polarity maintenance. The coordinated rewiring of the Rho GTPase network drives actin remodeling, ultimately enhancing migration. Notably, complete knockout of VPS39, leading to total autophagy collapse[14, 52], proves lethal due to irreversible mitochondrial disintegration (data not shown), indicating that this compensatory mechanism operates only under conditions of partial autophagic impairment. This finding not only reconciles contradictory conclusions of previous complete knockout studies but also offers a new mechanistic explanation for the accumulation of damage in cells with migration defects in neurodegenerative diseases[14, 53].

We performed 2-h long-term live-cell SIM imaging, during which the mitochondria transfer process appeared to last for approximately 1 h. This highlights the importance of extended imaging durations for uncovering long-term biological processes. These super-resolution imaging results reveal a spatiotemporal triage mechanism for mitochondrial quality control: when VPS39 deficiency cripples intracellular degradation, damaged mitochondria are spatially confined to the cell periphery—presumably as cargos for migrasome-mediated expulsion. Alternatively, some mitochondria are selectively transferred to adjacent cells via retraction fibers, establishing a previously unrecognized intercellular salvage pathway (Fig. 5e, Fig. S16). This elucidates how migrasomes function as cellular "safety valves", expelling damaged mitochondria into the extracellular space for intercellular clearance while facilitating the recycling of functional elements between cells. Through targeted organelle redistribution, migrasomes may mitigate autophagic impairment, thereby preserving tissue homeostasis.

In conclusion, our study demonstrates that VPS39 deficiency unleashes a migrasome-mediated quality control pathway during autophagy impairment using super-resolution microscopy. Mechanistically, this compensatory clearance is driven by RhoA/Rac1-dependent cytoskeletal remodeling and accelerated cell motility, which collectively orchestrate mitochondrial expulsion. This mechanism may represent a novel quality control pathway, resolves mitochondrial dysfunction by coordinating oxidative stress resolution and systemic homeostasis restoration. Its potential as a therapeutic target for autophagic obstruction-related pathologies, particularly in contexts such as neurodegenerative disorders where defective mitochondrial clearance is pathogenic, warrants further investigation.

**Methods**

**Chemicals and recombinant proteins**

Fibronectin (#K20438, OriLeaf), WGA-647 (#25559, Biolite), WGA-555 (#25539, Biolite), Lipofectamine 2000 (#11668-019, Thermo Fisher Scientific), Lipofectamine 3000 (#L3000015, Thermo Fisher Scientific), Chloroquine (C798394, Macklin), SAR407899 (#T7391, TargetMol), Cytochalasin B (C805410, Macklin), DAPI (#62248, Thermo Fisher Scientific), Lyso Tracker Green (#C1047S, Beyotime), WGA-595 (#25509, Biolite), WGA-647 (#25512, Biolite), Opti-MEM (Thermo Fisher Scientific), JC-1 (#C2006, Beyotime), Alexa Fluor 647-phalloidin (#A22287, Thermo Fisher Scientific), DCFDA/H2DCFDA Cellular ROS Assay Kit (#ab113851, Abcam), Live Cell Imaging Buffer (#A59688DJ, Thermo Fisher Scientific), trypan blue (#T6146, Sigma-Aldrich), HBSS (#BL561A, Biosharp), Trypan blue (#T8154, Sigma - Aldrich).

Anti-β-actin (#AA128, Beyotime), Anti-VPS39 (#ab224671, Abcam), anti-Lamp1 (#65596-1-MR, Proteintech Group), anti-Lamp2 (#66301-1-Ig, Proteintech Group), anti-RhoA (66733-1-Ig, Proteintech Group), anti-Rac1 (#66122-1-Ig, Proteintech Group), anti-ATG7 (#67341-1-Ig, Proteintech Group), anti-CDC42 (#10155-1-AP, Proteintech Group), anti-LC3-II (#14600-1-AP, Proteintech Group), anti-p62 (#29503-1-AP, Proteintech Group), Goat Anti-Rabbit ATTO-647N (#40839, Sigma-Aldrich), Goat anti-Mouse Alexa Fluor® 568 (#A-11011, Thermo Fisher Scientific), Goat anti-Rat Alexa Fluor® 488 (#ab150165, Abcam).

**Reagent treatment**

During CYB treatment, cells were incubated at a concentration of 5 μg/mL in MEM medium with 10% FBS for 1 h. For SAR treatment, cells were exposed to a 10 μM concentration in serum-free medium for 1 h. CQ treatment was performed at 50 μM in MEM medium with 10% FBS for 6 h; in the case of starvation treatment, cells were cultured in serum-free MEM medium for 4 h. Subsequently, the cells were washed three times with PBS before proceeding to the next step.

**Cell culture**

For cell lines, U2OS cells (human osteosarcoma cells) were cultured in 10% (v/v) FBS (fetal bovine serum; Gibco) and McCoy's 5A medium (Gibco). L929 cells were cultured in 10% (v/v) FBS (fetal bovine serum; Gibco) and MEM medium (Gibco). 293T cells were cultured in 10% (v/v) FBS (fetal bovine serum; Gibco) and DMEM medium (Gibco). All the cell culture medium contained 1% antibiotic-antimycotic (Gibco) and cells were maintained at 37°C with 5% $CO_2$ during culturing.

**Cell transfection and virus infection**

Approximately $1 \times 10^5$ of cells per well were seeded into a 24-well plate. After 24 h, transfection was performed with Opti-MEM medium and Lipofectamine 3000 according to the standard protocol. Cells were digested with trypsin 5 h after transfection and seeded onto cover.

Gene knockdown was accomplished using short hairpin RNA (shRNA) within the lentivirus-based vector pLKO.1-puro. Lentiviral production and infection procedures adhered to the manual guidelines. In brief, 5 μg of the target plasmid harboring the lentiviral vector, 10 μg of psPAX2, 10 μg of pMD2.G, and 50 μL of transfection reagent P3000 were amalgamated with 500 μL of Opti-MEM and allowed to stand at room temperature for 5 min. Subsequently, this mixture was combined with 35 μL of Lipo3000 transfection reagent in 500 μL of Opti-MEM and co-transfected into 293T cells. After 48 h, the supernatant was centrifuged at 1000 rpm for 3 min to eliminate cellular debris. The resultant viral particles were collected for subsequent experiments. For viral infection, the specified cells were seeded at 60%–70% confluence, with 8-10 μg/mL polybrene added to promote viral attachment to the cells. The cells were then incubated in fresh medium containing 5 μg/mL puromycin for selection until drug-resistant colonies were discernible.

**Construction of VPS39 downregulation and overexpression cell lines**

L929 cell lines with VPS39 knockdown and overexpression were established. Cells in the exponential growth phase were selected, and specific shRNA targeting VPS39 was designed and synthesized. According to the instructions, shRNA was transfected into L929 cells using liposome transfection reagent. After 72 h, WB was used to detect the expression level of VPS39 to verify the knockdown effect. In subsequent experiments, the knockdown cell lines were divided into shVPS39-1, shVPS39-2, and shVPS39-3 groups. A VPS39 overexpression plasmid was constructed. Once the cells reached the appropriate density, transfection was performed, and the plasmid was transfected into the cells using the corresponding transfection reagent. Stable expression cell lines were selected using resistance screening and the effect of VPS39 overexpression was verified by WB, with the overexpression cell line designated as the OE-VPS39 group.

**Confocal imaging and image analysis**

The confocal images were acquired with a Nikon C2 microscope, equipped with a 60x oil immersion objective with a numerical aperture (NA) of 1.45. An argon ion laser with a wavelength of 488 nm can excite the imaging of the green fluorescent protein GFP; a DPSS laser with a wavelength of 561 nm can be used as the imaging laser for the red fluorescent proteins RFP and mCherry. Imaging time is chosen based on cell conditions. Results were analyzed using Fiji, and data visualization was done with Origin.

**Structured Illumination Microscopy (SIM) Imaging**

SIM imaging was performed with a high-sensitivity structured illumination microscope (HiS-SIM, Guangzhou Computational Super-Resolution Biotech Co., Ltd., Guangzhou, China) equipped with 488, 560, and 638 nm excitation lasers and a 100×/1.49-NA oil-immersion objective. All the super resolution images were reconstructed from 9 raw images. SIM images were analyzed with Fiji software.

**Western blot (WB) analysis**

Protein expression was analyzed by WB. Cells were lysed in RIPA buffer supplemented with protease and phosphatase inhibitors. Protein concentrations were quantified using the BCA assay. Equal protein amounts (20-40 μg) were denatured in Laemmli buffer, separated by SDS-PAGE using 8-12% gels, and transferred onto PVDF membranes. The membranes were blocked with 5% non-fat milk or BSA in TBST for 1 h at room temperature and incubated overnight at 4°C with primary antibodies (diluted in blocking buffer). After three washes with TBST, the membranes were incubated with HRP-conjugated secondary antibodies for 1 h at room temperature. Protein bands were visualized using enhanced chemiluminescence (ECL) substrate and subsequently imaged. Band intensities were quantified using ImageJ software and normalized to β-actin or β-tubulin as loading controls.

**Intracellular reactive oxygen species (ROS) assay**

The DCFDA Cellular ROS Assay Kit was used to detect the accumulation of intracellular ROS in L929 cells. Briefly, cells were treated with 200 μL of 30 μM DCFDA working solution in serum-free medium at 37°C for 45 min in the dark. Fluorescence intensity was measured using a Nikon C2 microscope at Ex/Em. = 488/525 nm. The assay was conducted three independent times.

**Mitochondrial membrane potential assay**

Mitochondrial membrane potential (ΔΨm) was measured using the JC-1 probe with the mitochondrial membrane potential assay kit (Beyotime, China) according to the manufacturer's instructions. Briefly, L929 cells were seeded in 20 mm glass-bottom confocal dishes at a density achieving 60-70% confluence and cultured overnight at

37°C with 5% $CO_2$. Following PBS washing, cells were incubated with 10 μg/mL JC-1 working solution at 37°C with 5% $CO_2$ for 20 min, and then washed three times with HBSS buffer. Cells were imaged and quantitatively analyzed with Nikon C2 confocal microscope.

**Probe Labeling**

Live Cell Labeling: L929 or U2OS cells were seeded cells in 20 mm confocal dishes to reach a confluence of 50-70%. Discard the original medium, rinse cells with PBS, and add 200 μL of prepared probe working solution. Then incubate the cells at 37°C in a 5% $CO_2$ environment. For mitochondrial labeling, cells were incubated with 100 nM PK MITO Red in culture medium supplemented with 10% serum for 15 min. For migrasome labeling, cells were incubated with 5 μg/mL WGA-595 or WGA-647 in HBSS buffer for 30 min. For lysosomal staining, cells were incubated with 50 nM Lyso Tracker Green in serum-free medium, or with 5 μM Morph-SIR in serum-free medium for 30 min. Before imaging, a solution of Trypan blue (200 μL, 1 mg/mL) in PBS was added to exclude dead cells and quench the extracellular fluorescence from probes bound to either the cell membrane or the dish surface. After 1 min, Trypan blue was removed, and the cells were gently washed twice with PBS and immersed in Live Cell Imaging Buffer to optical imaging.

Fixing cells: The existing cell culture medium was discarded, and cells were washed three times with PBS. Subsequently, 200 μL of 4% paraformaldehyde (PFA; pre-warmed to 37°C) was added for fixation at room temperature for 30 min. After three 5-min washed with PBS, cells were incubated with a probe working solution for fixed cells. For actin filament labeling, cells were stained with 15 nM Phalloidin-647 in PBS for 30 min. The probe solution was then discarded, and cells were washed three times with 200 μL PBS (5 min per wash), followed by immersion in Live Cell Imaging Buffer for fluorescence imaging.

**Immunofluorescence**

Prior to experimentation, L929 cells were seeded in glass-bottom confocal dishes. After 24-h culture, cells reached approximately 60% confluence with uniform distribution and firm adhesion. Cells were washed three times with 200 μL PBS, fixed with 4% PFA in a 37°C incubator for 15 min, and subsequently washed three times with PBS. Blocking was performed at room temperature for 50 min using a solution of 5% goat serum and 0.2% Triton X-100 in PBS. Following blocking, cells were incubated overnight at 4°C with 200 μL of primary antibody dilution (1:200 in PBS). After primary antibody incubation, cells were subjected to three 5-min washes with PBS on an orbital shaker (100 rpm). Cells were then incubated with 200 μL of secondary antibody dilution (1:1000 in PBS) at room temperature for 1 h in the dark, followed by three additional 5-min PBS washes under shaking conditions (100 rpm). Finally, cells were preserved in fresh PBS protected from light pending subsequent imaging.

**Cell scratch assay**

L929 cells were seeded at a density of $5×10^5$ cells per well in 6-well plates and cultured overnight at 37°C in 5% $CO_2$ to achieve >90% confluence. Perpendicular scratch wounds were generated on the confluent monolayer using ethanol-sterilized 200-μL pipette tips with gentle pressure. After washing with PBS 1-2 times to remove detached cells, images were acquired at designated time points (0, 12, and 24 h post-scratch) using an Olympus CKX53 microscope. Eight or more fields per sample were selected for analysis with triplicate independent experiments per group. Finally, we use ImageJ software to measure the scratch area ratio and calculate the cell healing area percentage, or migration rate, as follows: [(initial scratch area - scratch area at any time point) / initial scratch area] × 100%.

**Statistical analyses**

Data in graphs are presented as means ± SEM. Statistical analyses were performed using SPSS. The statistical differences between experimental and control groups were assessed analyzed by t test analysis. Differences between groups were assessed using analysis of variance followed by a two-tailed unpaired Student's t-test. Statistical significance was defined as $P < 0.05$. Graphical representations of data were generated using Origin.

**Author contributions**

Y.H. conceived of the project. Experiments were performed primarily by X.P. and W.S. with regular input and guidance from Y.H., Y.-H.Z. and H.L.. N.J. created the mechanism diagrams, W.G. contributed to the guidance of the experiments. Y.H. and X.P. wrote the manuscript with discussion and improvements from all authors.

**Competing interests**

The authors declare no competing financial interests.

**Acknowledgment**

We thank the Optical Bioimaging Core Facility of WNLO-HUST (Wuhan National Laboratory for Optoelectronics-Huazhong University of Science and Technology), the Research Core Facilities for Life Science of HUST, and Advanced Biomedical Imaging Facility-WNLO for the support in data acquisition. This work was supported by the following grants: National Natural Science Foundation of China (grant nos. 92354305, 32271428, 32201132), STI 2030—Major Projects (2021ZD0200401), Zhejiang Provincial Natural Science Foundation of China (LZ25F050008, LZ24H310002), Natural Science Foundation of Hangzhou (2024SZRZDH160001).

**Data availability**

The authors declare that all data supporting the findings of this study are available within the article and its Supplementary Information files or from the corresponding authors on reasonable request.

**Supporting information**

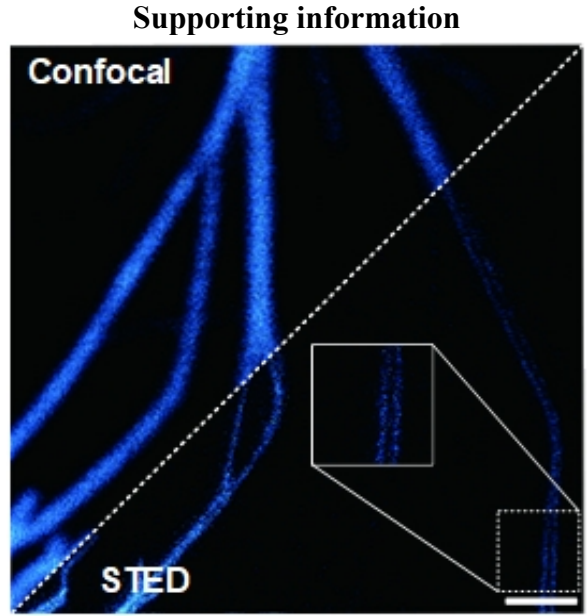

**Fig S1 Comparative confocal and STED imaging of migrasomes.** Confocal and STED images of migrasomes in L929 cells labeled with WGA-595. Scale bars: 2 μm.

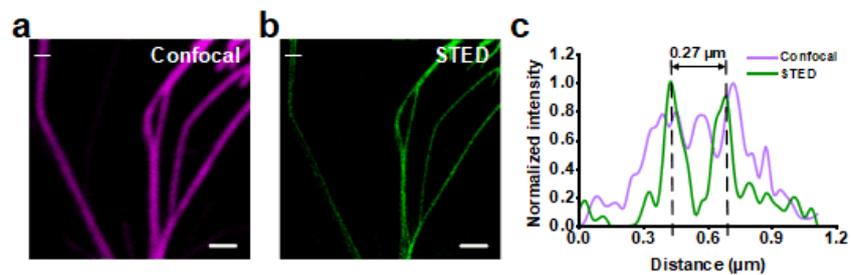

**Fig S2 STED microscopy resolves the luminal structures of retraction fibers undetected by confocal microscopy. a, b** Representative confocal (a) and STED (b) images of WGA-595-labeled L929 cell migrasomes. **c** Normalized fluorescence intensity distribution along the line in (a) and (b). Scale bars: 2 μm.

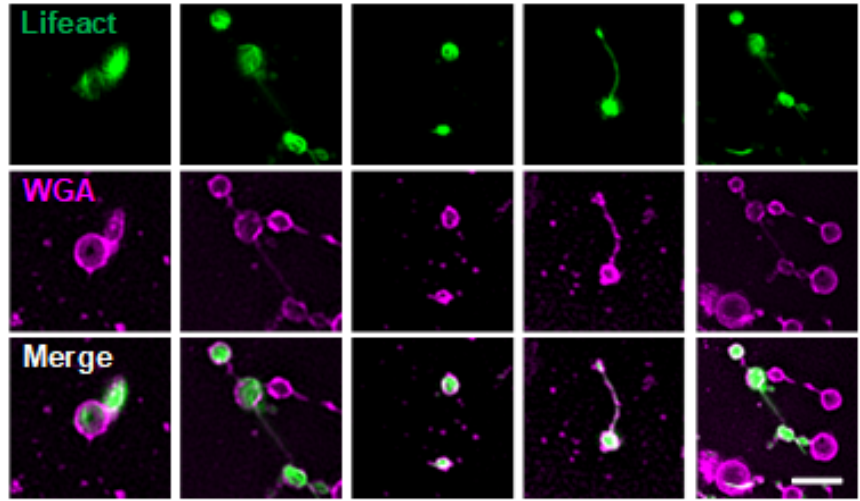

**Fig S3 Colocalization images of migrasomes and actin filaments under Super-resolution microscope.** SIM images of migrasomes (WGA-647, magenta) and actin filaments (Lifeact-EGFP, green) in live L929 cells. Scale bar: 2 μm.

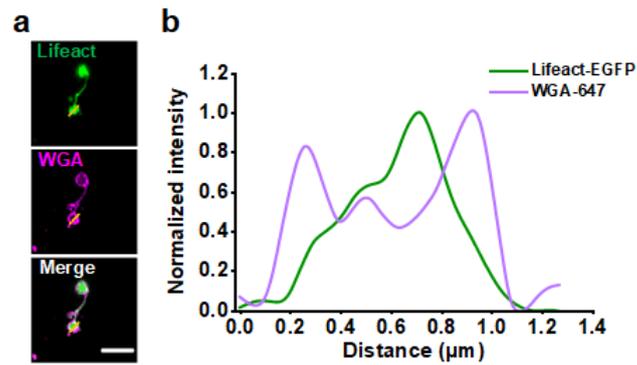

**Fig S4 Super-resolution imaging reveals actin filaments encapsulated within migrasomes. a** SIM images of migrasomes (WGA-647, magenta) and actin filaments (Lifeact-EGFP, green) in live L929 cells. **b** Normalized fluorescence intensity distribution along the line in (a). Scale bar: 2 μm.

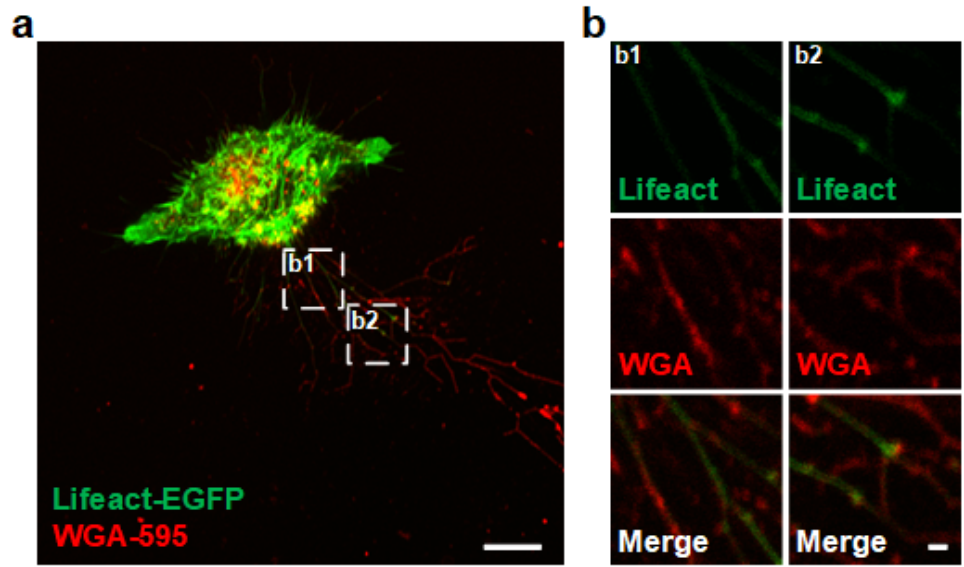

**Fig S5 Colocalization analysis of migrasomes and actin filaments under confocal microscope.** a Confocal imaging of migrasomes (WGA-595, red) and actin filaments (Lifeact-EGFP, green) in live L929 cells. b Enlargements of the boxed regions in (a). Scale bar:10 μm (a), 1 μm (b).

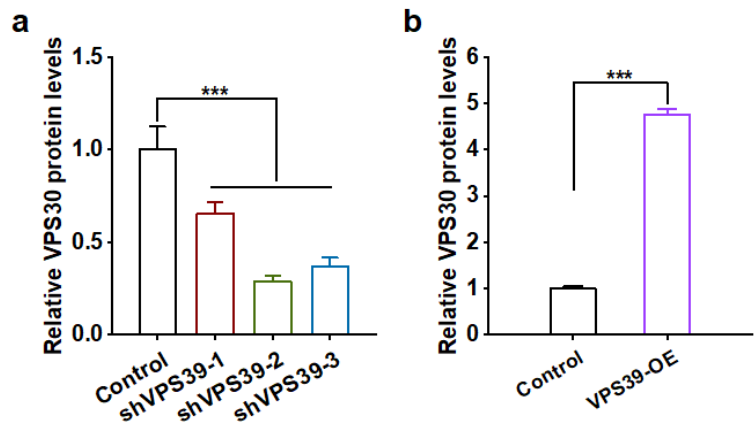

**Fig S6 Evaluation of VPS39 protein levels in stable VPS39 Knockdown and overexpression L929 cell lines. a, b** Quantification of relative VPS39 protein levels. Data in (a, b) normalized to control, represent mean ± SEM, *$p < 0.05$, **$p < 0.01$, ***$p < 0.001$, n.s. not significant ($p > 0.05$).

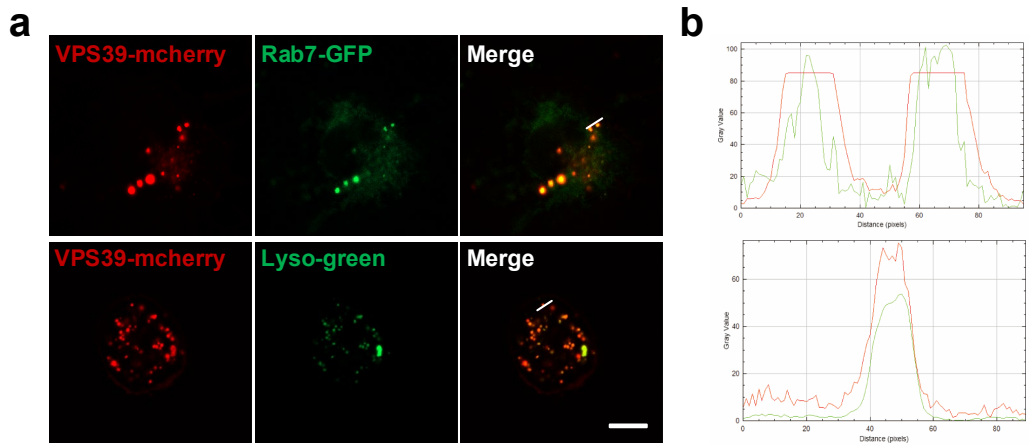

**Fig S7 Colocalization analysis of VPS39 with Rab7 and lysosomes. a** Confocal images of L929 cells expressing VPS39-mCherry (red) and Rab7-GFP (green). The merged image demonstrates overlapping signals (yellow) indicating spatial association. **b** Fluorescence intensity distribution along the line in (a). Scale bar: 10 μm.

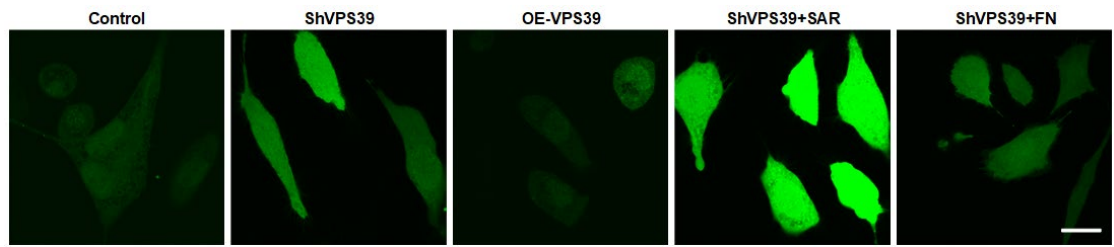

**Fig S8 VPS39 regulates intracellular ROS production.** Confocal imaging was employed to measure intracellular ROS levels (DCFDA fluorescence intensity) in L929 cells under indicated conditions. Scale bar: 10 μm.

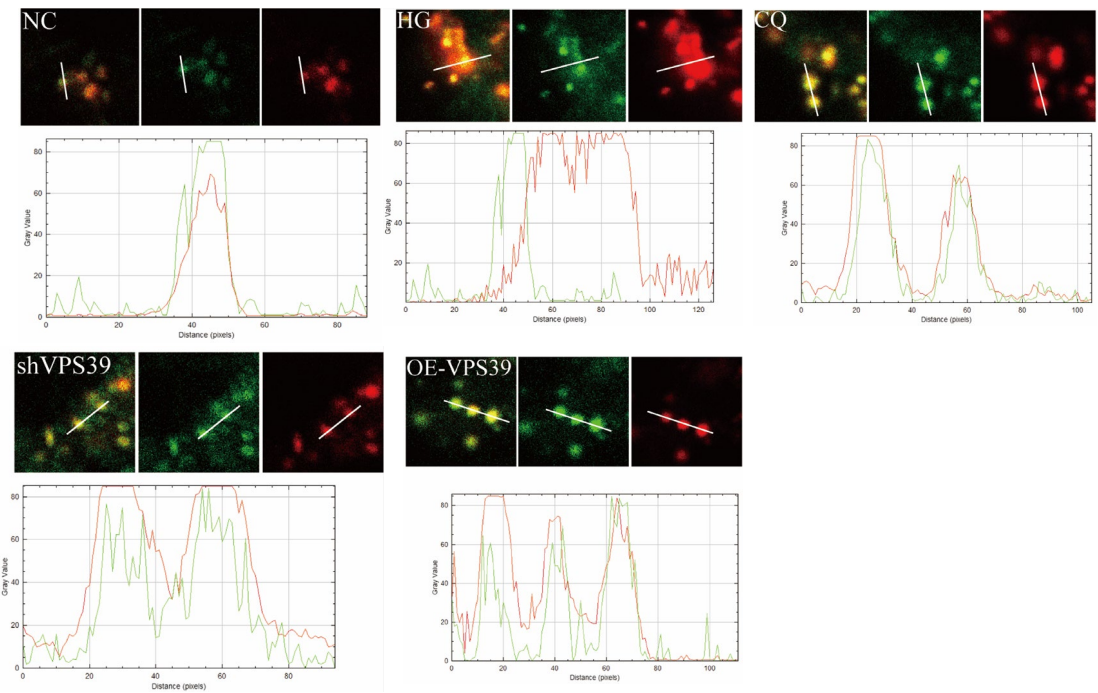

Fig S9 Confocal images of L929 cells expressing the autophagic flux sensor GFP-LC3-RFP-LC3ΔG under indicate conditions, and fluorescence intensity distribution along the line.

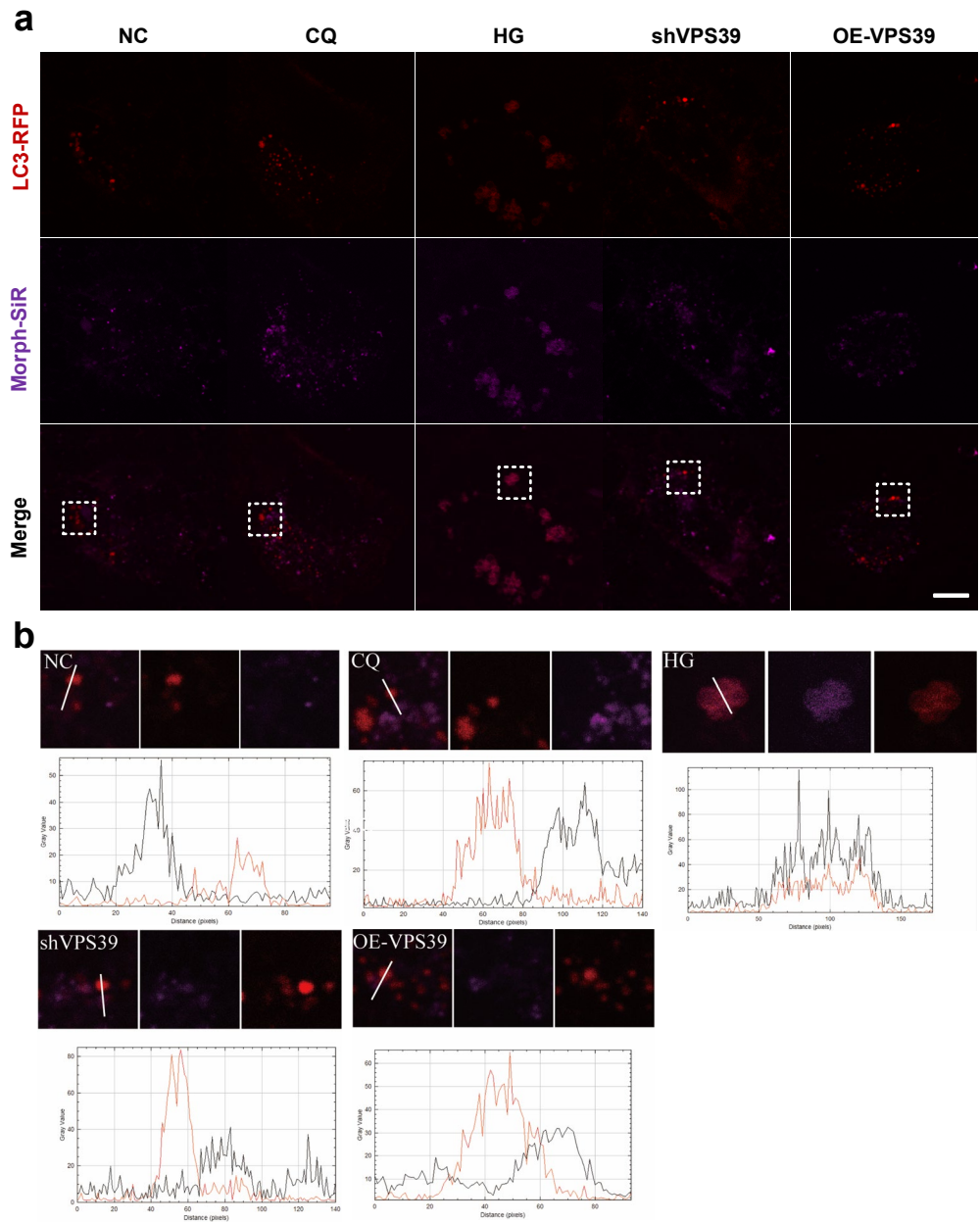

**Fig S10 VPS39 knockdown impairs autophagic flux and modifies lysosomal structure. a** Confocal images of L929 cells expressing the autophagic flux sensor GFP-LC3-RFP-LC3ΔG (autolysosomes, red) stained with Morph-SiR (Lysosomal probe, magenta) under indicated conditions. Scale bars: 10 μm. **b** The magnified area of the region of interest (ROI) in (a), accompanied by the distribution of fluorescence intensity along the line under indicated conditions.

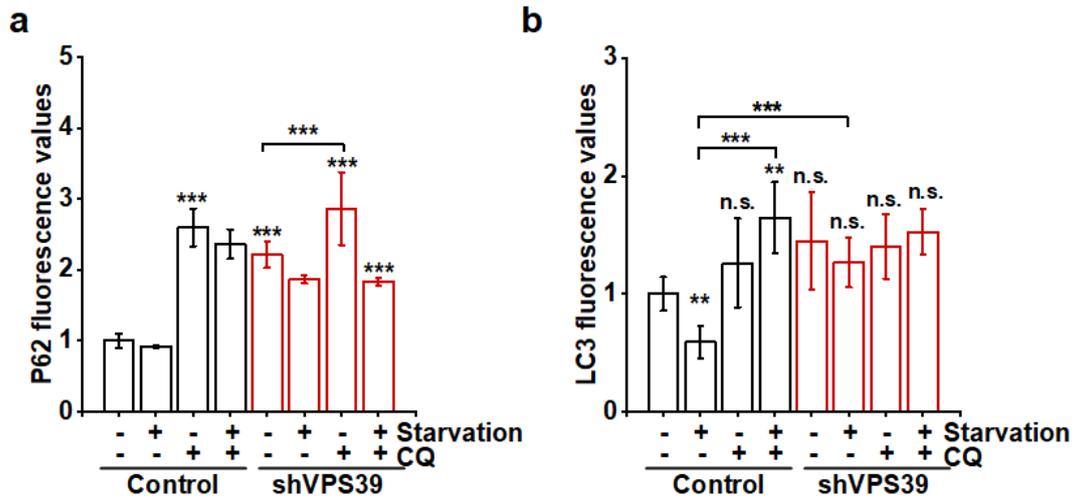

**Fig S11 Quantitative analysis of endogenous p62 and LC3-II immunofluorescence intensity under various cellular conditions. a, b** Fluorescence intensity quantification of endogenous p62 (a) and LC3-II (b) under indicated conditions. Data in (a, b) normalized to control, represent mean ± SEM, *$p < 0.05$, **$p < 0.01$, ***$p < 0.001$, n.s. not significant ($p > 0.05$).

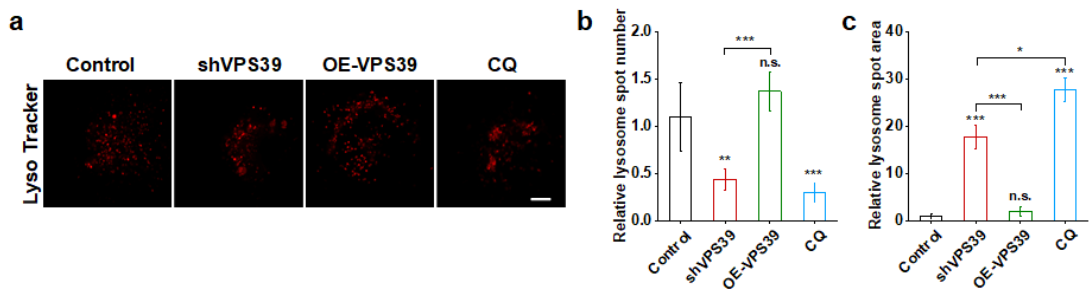

**Fig S12 VPS39 depletion or lysosomal dysfunction alters lysosomal biogenesis. a** Confocal images of lysosomes stained with Lyso Tracker Green (red) in Control, shVPS39, OE-VPS39, and CQ-treated L929 cells. Scale bars: 10 μm. **b, c** Quantification of lysosomal puncta number (b) and area (c) per L929 cell. Data are presented as mean ± SEM, *$p < 0.05$, **$p < 0.01$, ***$p < 0.001$, n.s. not significant ($p > 0.05$).

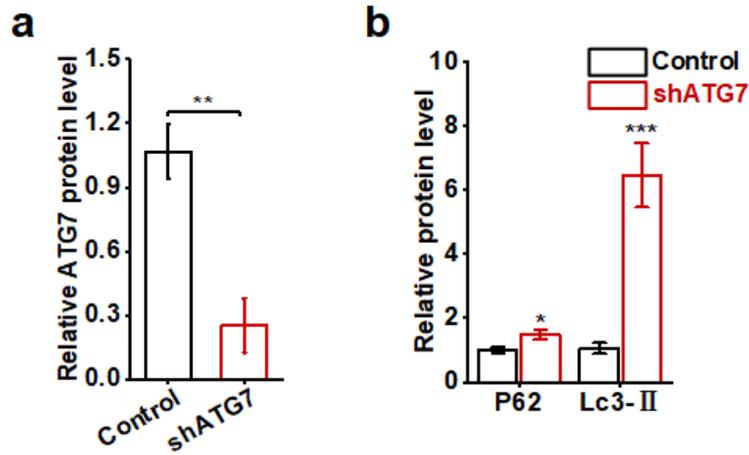

**Fig S13 ATG7 knockdown dysregulates core autophagy proteins. a, b** Quantitative analysis of ATG7, P62, and LC3-II protein expression in control and ATG7-knockdown (shATG7). Data are presented as mean ± SD, *p < 0.05, **p < 0.01, ***p < 0.001, n.s. not significant (p > 0.05).

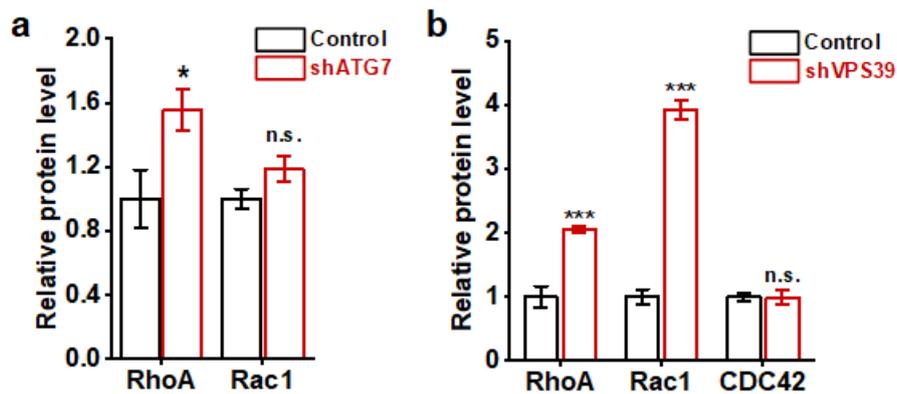

**Fig S14 ATG7 or VPS39 depletion differentially regulates Rho GTPase expression. a.** Quantitative analysis of protein RhoA and Rac1 under control and shATG7. **b.** Quantitative analysis of protein RhoA, Rac1, and CDC42 under control and VPS39 knockdown (shVPS39). Data are presented as mean ± SD, *p < 0.05, **p < 0.01, ***p < 0.001, n.s. not significant (p > 0.05).

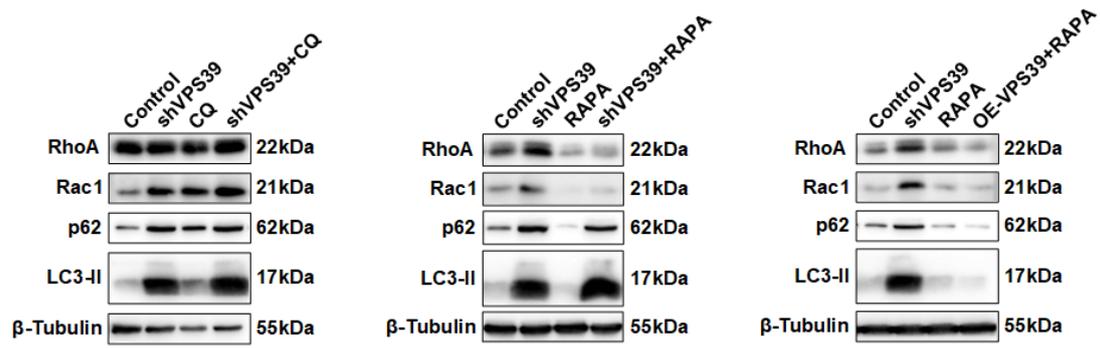

**Fig S15 Effects of VPS39 Modulation on RhoA, Rac1, and Autophagy Markers Under Autophagy-Modulating Conditions.** Western blot of RhoA, Rac1, P62 and LC3-II) under indicated conditions: control, shVPS39, chloroquine (CQ) treatment, CQ treatment following shVPS39, rapamycin (RAPA) treatment, RAPA treatment following shVPS39, and after RAPA treatment subsequent to VPS39 overexpression (OE-VPS39). β-tubulin: loading control.

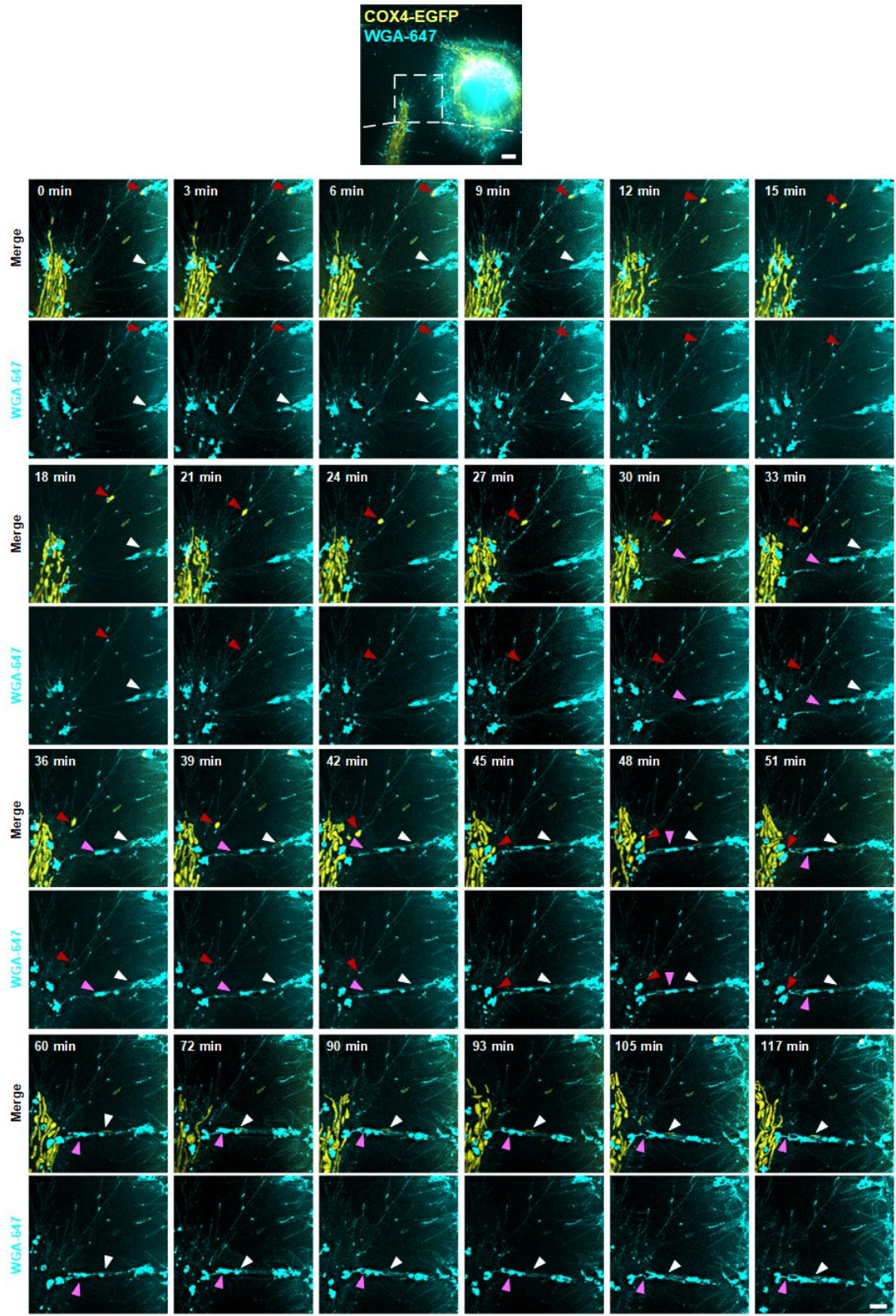

**Fig S16 Super-resolution time-lapse imaging captures the intercellular transfer of mitochondria in living cells.** Magnified time lapse SIM imaging (3-min intervals) of mitochondria (COX4-EGFP, green) and migrasomes (WGA-647, magenta) in live L929 cells. Bottom: Magnified ROIs. Scale bars: 5 μm (main panels), 2 μm (magnified panels).